\documentclass[useAMS]{mn2e}

\newcommand{\Msolar}{\mbox{\,$\rm M_{\odot}$}} 
\newcommand{\Lsolar}{\mbox{\,$\rm L_{\odot}$}} 

\usepackage{graphicx}
\usepackage{subfigure}

\title[The Pulsation of AGB Stars in NGC\,1978 and NGC\,419]
{The Pulsation of AGB Stars in the Magellanic Cloud Clusters NGC\,1978 
and NGC\,419}
\author[D. Kamath et al.]{D. Kamath$^{1}$\thanks{E-mail: 
devika13@mso.anu.edu.au; 
wood@mso.anu.edu.au; soszynsk@.astrouw.edu.pl; 
thomas.lebzelter@univie.ac.at}, 
P.\,R. Wood$^{1}$\footnotemark[1], I. Soszy\'nski$^{2}$\footnotemark[1] 
and T. Lebzelter$^{3}$\footnotemark[1]\\
$^{1}$Research School of Astronomy \& Astrophysics, Australian National 
University, Weston Creek, ACT 2611, Australia\\
$^{2}$Warsaw University Observatory, Aleje Ujazdowskie 4, 00-478 Warszawa, 
Poland\\
$^{3}$Institute of Astronomy, University of Vienna, Tuerkenschanzstrasse 17, 
1180 Vienna, Austria}

\begin{document}

\date{Accepted . Received ; in original form }

\pagerange{\pageref{firstpage}--\pageref{lastpage}} \pubyear{2010}

\maketitle

\label{firstpage}

\begin{abstract}

The intermediate-age Magellanic Cloud clusters NGC\,1978 and NGC\,419
are each found to contain substantial numbers of pulsating AGB stars,
both oxygen-rich and carbon-rich.  Each cluster also contains two
pulsating asymptotic giant branch (AGB) stars which are infrared
sources with a large mass loss rate.  Pulsation masses have been
derived for the AGB variables, from the lowest luminosity O-rich
variables to the most evolved infrared sources.  It is found that the
stars in NGC\,1978 have a mass of 1.55\Msolar~early on the AGB while
the NGC\,419 stars have a mass of 1.87\Msolar~early on the AGB.  These
masses are in good agreement with those expected from the cluster ages
determined by main-sequence turnoff fitting.
Nonlinear pulsation models fitted to the highly evolved AGB stars show
that a substantial amount of mass loss has occurred during the AGB
evolution of these stars.  An examination of the observed mass
loss on the AGB, and the AGB tip luminosities, shows that in both
clusters the mass loss rates computed from the formula of Vassiliadis
\& Wood (1993) reproduce the observations reasonably well.  The mass
loss rates computed from the formula of Bl\"ocker (1995) terminate the
AGB in both clusters at a luminosity which is much too low.

\end{abstract}

\begin{keywords}
stars: AGB and post-AGB -- stars: mass loss -- stars: variables: 
general -- galaxies: Magellanic clouds -- galaxies: star clusters : 
NGC\,1978, NGC\,419.
\end{keywords}

\section{Introduction}

Star clusters are ideal sites to test the theories of 
stellar evolution, mass loss and pulsation. They contain a rather 
homogeneous sample of stars with well constrained parameters 
such as mass, metallicity, age, distance and luminosity. Furthermore, 
the star clusters associated with the Magellanic Clouds span a wide 
range of age, which enables us to study the evolution of stars 
with various masses. Owing to the relatively large numbers of stars 
in many of the star clusters in the Magellanic Clouds they are suitable for 
studying the short-lived later stages of stellar evolution. In 
this paper we target two intermediate age clusters, the Large 
Magellanic Cloud (LMC) cluster NGC\,1978 and the Small Magellanic 
Cluster (SMC) cluster NGC\,419. These two clusters are the only 
ones in the Magellanic Clouds which are known to have AGB stars 
with extremely high mass loss rates. Each cluster has one star 
that was first detected by the ISOCAM survey as a mid-IR source 
(Tanab\'e et al. 1998) as well as one slightly less extreme star 
found first in the near-IR (Tanab\'e et al. 1997). The clusters 
also have many other less extreme, but pulsating, AGB stars. 

Our aim here is to use the pulsation properties of the 
AGB stars in these clusters to derive accurate mass estimates for the 
cluster AGB stars and to thereby study mass loss along the AGB. This mass loss 
along the AGB can then be compared with that expected from mass 
loss laws commonly used in stellar evolution calculations. 

The cluster NGC\,1978 is a massive ($\sim$$2\times10^{5}$\Msolar, 
Westerlund 1997) and luminous red globular cluster in the LMC. It has 
a rather high ellipticity (${\epsilon} = 0.3$, Fischer et al. 1992; 
Mucciarelli et al. 2007) suggesting it may be the result of the 
merger of two clusters. The cluster is found to contain oxygen-rich 
(M-type) stars with C/O $<$ 1 in the stellar atmosphere and 
carbon-rich (C-type) stars with C/O $>$ 1 (e.g. Frogel et al. 1990). 
Ferraro et al. (2006) derived a cluster metal abundance [Fe/H] = -0.38 
dex and noted that there was no evidence for a significant dispersion in 
abundance between stars ($\sigma = 0.07$ dex). Olszewski et al. (1991) 
derived [Fe/H] = -0.42$\pm$0.04 dex. However, Hill et al. (2000) 
derived a significantly lower abundance ([Fe/H] = -0.96$\pm$0.15) dex. 
This value has not been reproduced by any other group (we will assume 
[Fe/H] = -0.4, corresponding to Z $\approx$ 0.008, in our 
pulsation models). In a recent work, Mucciarelli et al. (2007) used 
Hubble Space Telescope (HST) observations, along with a distance 
modulus to the LMC of 18.50 and reddening of E(B-V) = 0.09, to 
determine the age of this cluster to be $\tau = 1.9\pm0.1$ 
Gyr when using isochrones from Pisa Evolutionary Library 
(Castellani et al. 2003) with $Z = 0.008$, consistent with 
the determinations of the cluster metallicity. Bomans, Vallenari \& 
de Boer (1995) derived an age of 2.2 Gyr for NGC\,1978, and 
found no evidence for age variation within the cluster (they 
assumed a distance modulus of 18.50, E(B-V) = 0.08 and a 
metallicity of [Fe/H] = -0.4). The homogeneous nature of the isochrone 
of NGC\,1978 is confirmed by the HST study of Milone et al. (2009). 
The age determined for NGC\,1978 (1.9 to 2.2 Gyr) corresponds to 
an initial mass for stars at the beginning of the thermally 
pulsing AGB (TPAGB) of 1.54 to 1.62 \Msolar~according to the 
isochrones of Girardi et al. (2000) with metallicity $Z = 0.008$. 
These isochrones were derived from stellar evolution models that 
included a scaled Reimers mass loss law. The mass estimated at 
the beginning of the TPAGB for stars of age 1.9 to 2.2 Gyr is 
1.44 to 1.53 \Msolar.

The cluster NGC\,419 is one of the brightest intermediate age globular
clusters in the SMC. Recent deep HR-diagrams obtained with the HST
show that NGC\,419 has a broad main-sequence turnoff which is
interpreted as the result of multiple star formation events or
extended star formation (Glatt et al. 2008; Girardi et al. 2009;
Rubele et al. 2009).  The metallicity of NGC\,419 was determined from
calcium triplet of observations by Kayser et al. (2009) to be close to
[Fe/H] = -0.7, corresponding to a metallicity of $Z \approx
0.004$. Fits to the HR-diagrams yield ages of 1.2--1.6 Gyr (Glatt et
al. 2008), 1.3--1.5 Gyr (Girardi et al. 2009) or 1.2--1.8 Gyr (Rubele
et al. 2009).  If we adopt ages of 1.4$\pm$0.2 Gyr, then the
isochrones of Girardi et al. (2000) with a metallicity of $Z = 0.004$
yield an initial mass for stars at the beginning of the TPAGB of
1.82$\pm$0.15 \Msolar~and a current mass (as a result of mass loss via
a scaled Reimers law) of 1.79$\pm$0.15 \Msolar.  The two red giant
clumps observed in NGC\,419 were shown by Girardi et al. (2009) to be
the result of the mass range of red giant stars in the cluster
spanning the critical mass ($\approx$1.9 \Msolar) which separates
those stars that do, or do not, develop electron-degenerate cores on
entering the red giant branch. The distance moduli derived by Glatt et
al. (2008) and Rubele et al. (2009) are 18.83 and 18.84, respectively,
and both estimated the reddening as E(B-V) = 0.11.

\section[]{Observational Data}

\begin{table*}
 \begin{minipage}{170mm}
\centering
  \caption{AGB variables of NGC\,1978.  The RA and DEC are the 2MASS coordinates. 
The $B$ and $R$ band magnitudes are from MACHO. The other other magnitudes 
are from our near-IR photometry.  The last column gives the spectral type.}
  \begin{tabular}{@{}llccccccccc}
 \hline
   Ident & MACHO & \multicolumn{2}{c}{RA (J2000) DEC} & $<$$B$$>$ & $<$$R$$>$ & $<$$J$$>$ & $<$$H$$>$ & $<$$K$$>$ & $<$$L$$>$ & SpT\\
  \hline
LE3  & 64.7844.15  & 5 28 44.4 & -66 14 04 & 16.40 & 14.21 & 11.08 & 10.34 &  9.70 &  8.99 & C \\
LE6  & 64.7844.16  & 5 28 46.2 & -66 13 56 & 15.80 & 14.33 & 12.21 & 11.17 & 10.81 & 10.37 & C \\
LE4  & 64.7844.18  & 5 28 43.7 & -66 14 04 & 15.93 & 14.54 & 12.49 & 11.58 & 11.34 & 11.21 & M \\
B    & 64.7844.20  & 5 28 43.6 & -66 14 09 & 16.24 & 14.74 & 12.38 & 11.45 & 11.17 & 10.89 & C \\
LE5  & 64.7844.23  & 5 28 43.5 & -66 13 53 & 16.19 & 14.92 & 12.57 & 11.67 & 11.47 & 11.31 & M \\
IR1  & 64.7844.416 & 5 28 40.2 & -66 13 54 &  0.00 & 17.47 & 14.93 & 13.03 & 11.16 &  9.01 & C \\
A    & 64.7845.14  & 5 28 46.1 & -66 13 25 & 15.54 & 14.13 & 12.25 & 11.34 & 11.14 & 10.97 & M \\
LE11 & 64.7845.20  & 5 28 43.6 & -66 14 09 & 16.44 & 14.83 & 12.84 & 11.95 & 11.56 & 11.09 & C \\
LE1  & 64.7965.15  & 5 28 48.4 & -66 15 00 & 16.62 & 14.75 & 12.13 & 10.97 & 10.35 &  0.00 & C \\
LE8  & 64.7965.16  & 5 28 48.4 & -66 14 39 & 16.12 & 14.82 & 13.02 & 12.15 & 11.95 & 11.73 & M \\ 
LE7  & 64.7965.22  & 5 28 47.8 & -66 14 44 & 16.55 & 14.83 & 12.54 & 11.50 & 11.16 & 10.67 & C \\
LE2  & 64.7965.26  & 5 28 48.5 & -66 15 18 & 17.12 & 15.11 & 12.87 & 11.69 & 11.20 &  0.00 & C \\
MIR1 &             & 5 28 47.1 & -66 14 14 &  0.00 &  0.00 & 15.96 & 14.63 & 12.67 &  9.77 & C \\
 \hline
 \end{tabular}
 \end{minipage}
\end{table*}

In order to look for periodic variations of the AGB variables in the two
clusters NGC\,1978 and NGC\,419, the light curves of stars within 1.5 arcminutes of cluster centers were
analysed. For the LMC cluster NGC\,1978, data were extracted from the Massive
Compact Halo Object (MACHO) survey (B, R bands) (Alcock et al. 1997) and twelve AGB
variables were found. This does not include the mid-IR source NGC\,1978 MIR1 which was
not detected by the MACHO observations.
Nineteen AGB variables were found in the SMC cluster NGC\,419 using
light curve data from the Optical Gravitational Lensing Experiment OGLE II 
(V, I bands) database (\.Zebru\'n et al. 2001), supplemented with some 
OGLE III data.  The mid-IR source NGC\,419 MIR1 was not detected by the OGLE
observations.

Figures~\ref{fig:n1978LC} and \ref{fig:n419LC} show the MACHO and OGLE
light curves of the variables in NGC\,1978 and NGC\,419, respectively, while
Table 1 and Table 2 provide parameters of the variables.
In these tables, the candidates with LE numbers were identified 
with the AGB stars found by Lloyd Evans (1980). The near-IR (IR1) and the mid-IR (MIR1)
variables in both the clusters are the two very red stars with an IR excess. 
The two stars A and B in NGC\,1978 follow the naming from Lederer et al (2009). 
The star ADQR1 in NGC\,419 comes from Azzopardi et al. (1986). 

In order to determine periods for the two MIR sources, as well as to
obtain accurate mean near-IR magnitudes for all the variables, the two
clusters were monitored relatively sparsely in time over approximately
3000 days with the 2.3m telescope of the Australian National
University at Siding Spring Observatory. The observations were taken
using the near-IR imaging system Cryogenic Array Spectrometer/Imager
(CASPIR) (McGregor et al. 1994). The observations were carried out
using the filters $J$ (1.28$\mu$m), $H$ (1.68$\mu$m), $K$ (2.22$\mu$m), and
$L$ (3.59$\mu$m). Mean flux-averaged photometric magnitudes in the
MACHO, OGLE and near-IR bandpasses are tabulated in Table 1 and Table
2 for the AGB variables in NGC\,1978 and NGC\,419, respectively.  The
$K$ and $L$ light curves of the near-IR and mid-IR sources are shown in
Figures~\ref{fig:IR1_MIR1_1978} and \ref{fig:IR1_MIR1_419}. 

In order to obtain the complete spectral energy distribution, the
Spitzer Space Telescope (SST) Surveys, SAGE of the LMC (Meixner et
al. 2006) and S$^{\rmn{3}}$MC of the SMC (Bolatto et al. 2007) were
also used to obtain photometric data covering the IRAC (3.6, 4.5, 5.8
and, 8 $\mu$m) bands.  Some of the stars in these two clusters were
observed by Groenewegen et al. (2007) with the SST infrared
spectrograph (IRS) from $\sim$5--37 $\mu$m.  For these stars the IRS
spectra were used instead of the IRAC data in the mid-IR.

Spectral types for some of the AGB variables in our study were
taken from Frogel et al. (1990). For the AGB variables without a
known spectral type, spectra were obtained with the Dual Beam
Spectrograph on the 2.3m telescope of the Australian National
University at Siding Spring Observatory. The spectral types for the
AGB variables in the clusters NGC\,1978 and NGC\,419 are listed in
Table 1 and Table 2 respectively.  The two mid-IR sources NGC\,1978 MIR1 and
NGC\,419 MIR1 are assumed to be of spectral type C since their
mid-IR spectra show features of SiC (Groenewegen et al. 2007).

\begin{table*}
 \begin{minipage}{170mm}
 \centering
  \caption{The AGB variables of NGC\,419. 
The $B$, $V$ and $I$ band magnitudes are from OGLE. The other magnitudes 
are from our near-IR photometry. The last column gives the spectral type.}
  \begin{tabular}{@{}lccccccccc}
  \hline
   Ident & OGLE & $<$$B$$>$ & $<$$V$$>$ & $<$$I$$>$ & $<$$J$$>$ & $<$$H$$>$ & $<$$K$$>$ & $<$$L$$>$ & SpT\\  
  \hline
LE16  & OGLE\,010801.10-725317.1 &  0.00 & 19.50 & 15.58 & 13.96 & 12.53 & 11.08 &  0.00 & C \\
ADQR1 & OGLE\,010810.31-725307.6 &  0.00 & 17.57 & 15.20 & 13.87 & 12.88 & 12.56 & 12.14 & C \\
LE20  & OGLE\,010811.55-725314.7 &  0.00 & 17.15 & 14.25 & 12.56 & 11.50 & 11.00 & 10.45 & C \\
LE21  & OGLE\,010812.36-725315.5 &  0.00 & 17.27 & 14.34 & 12.67 & 11.56 & 10.90 & 10.38 & C \\ 
IR1   & OGLE\,010812.92-725243.7 &  0.00 &  0.00 & 16.13 & 13.56 & 11.74 & 10.80 &  9.21 & C \\
5-3   & OGLE\,010814.58-725356.7 & 17.89 & 16.19 & 14.21 & 12.91 & 12.14 & 11.95 &  0.00 & M \\ 
LE25  & OGLE\,010815.63-725251.6 &  0.00 &  0.00 & 14.03 & 12.59 & 11.58 & 11.11 & 10.76 & C \\ 
LE24  & OGLE\,010815.73-725254.3 &  0.00 & 16.26 & 14.28 & 13.15 & 12.20 & 12.02 & 11.97 & -- \\ 
LE35  & OGLE\,010817.45-725301.1 &  0.00 & 17.30 & 14.34 & 12.52 & 11.49 & 10.86 & 10.30 & C \\ 
LE37  & OGLE\,010819.40-725312.0 &  0.00 & 16.40 & 14.27 & 12.77 & 11.85 & 11.56 & 11.33 & C \\ 
LE36  & OGLE\,010819.80-725259.0 &  0.00 & 17.47 & 14.33 & 12.55 & 11.17 & 10.71 & 10.28 & C \\ 
LE27  & OGLE\,010820.61-725251.7 & 19.11 & 17.17 & 14.30 & 12.52 & 11.44 & 10.89 & 10.46 & C \\ 
LE28  & OGLE\,010821.50-725216.0 & 19.28 & 17.29 & 14.89 & 13.23 & 12.02 & 11.53 & 10.85 & C \\ 
LE23  & OGLE\,010821.67-725302.7 &  0.00 & 16.87 & 14.29 & 12.83 & 11.72 & 11.23 & 10.82 & C \\ 
LE22  & OGLE\,010822.23-725302.3 &  0.00 & 17.48 & 14.52 & 12.78 & 11.69 & 11.06 & 10.64 & C \\ 
LE29  & OGLE\,010822.28-725233.5 & 19.57 & 16.91 & 14.30 & 12.74 & 11.66 & 11.24 & 10.75 & C \\
LE19  & OGLE\,010823.43-725318.2 & 18.00 & 16.34 & 14.55 & 13.38 & 12.52 & 12.34 & 12.14 & M \\ 
5-15  & OGLE\,010823.80-725309.4 & 18.10 & 16.35 & 14.54 & 13.39 & 12.53 & 12.33 & 11.94 & M \\ 
LE18  & OGLE\,010824.89-725256.7 &  0.00 & 18.17 & 14.70 & 12.82 & 11.60 & 10.93 & 10.21 & C \\ 
MIR1  & MIR~~~010817.47-725309.5 &  0.00 &  0.00 &  0.00 &  0.00 &  0.00 & 14.64 & 10.75 & C \\ 
 \hline
 \end{tabular}
 \end{minipage}
\end{table*}

\section[]{Light Curves and Pulsation Periods}

\begin{figure*}
\centering \includegraphics[width=150mm]{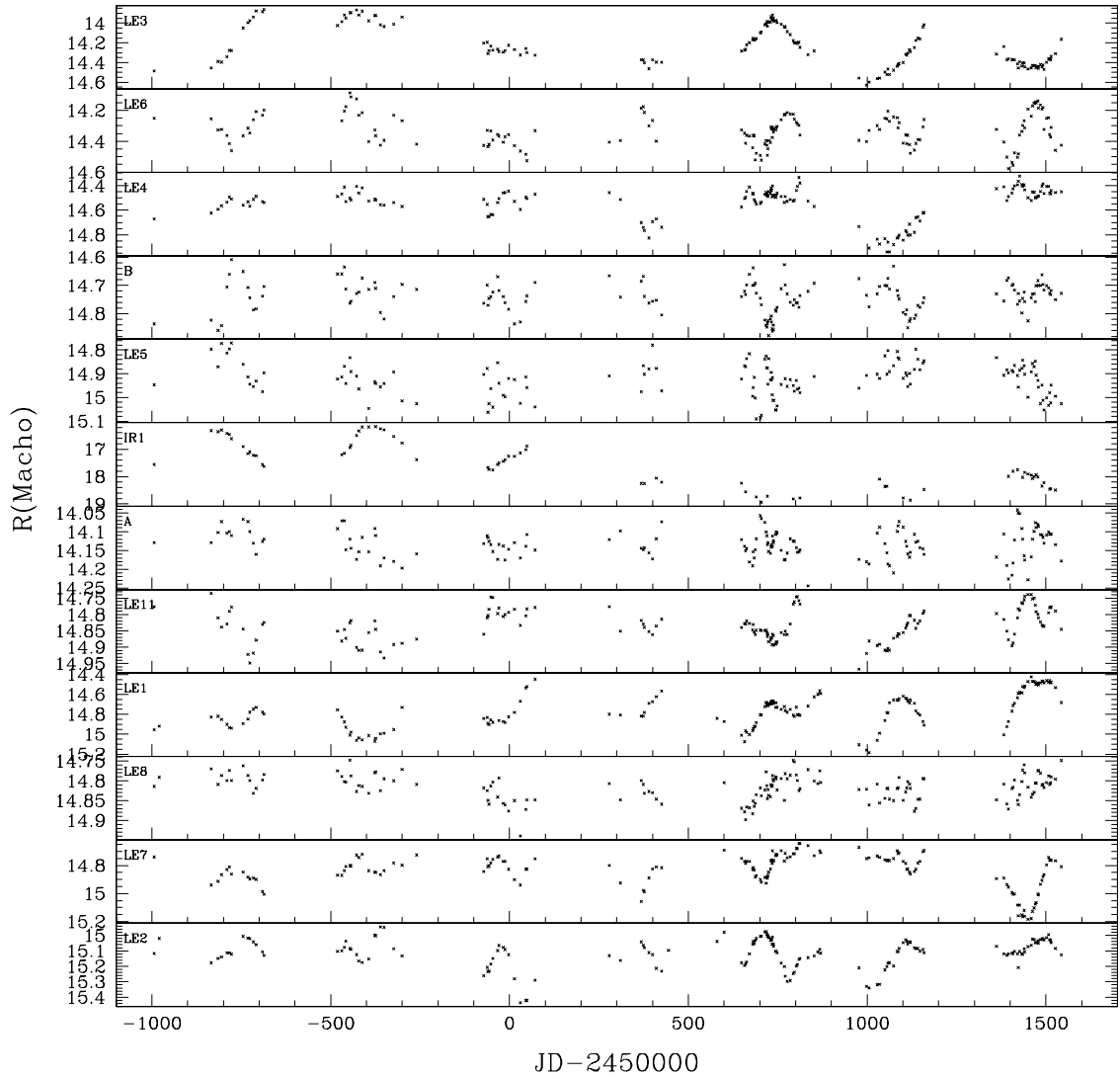}
 \caption{MACHO $R$ band light curves of the variable AGB stars in NGC\,1978.}
\label{fig:n1978LC}
\end{figure*}

\begin{figure*}
\begin{center}
\begin{minipage}{200pt}
\includegraphics[width=2.75in]{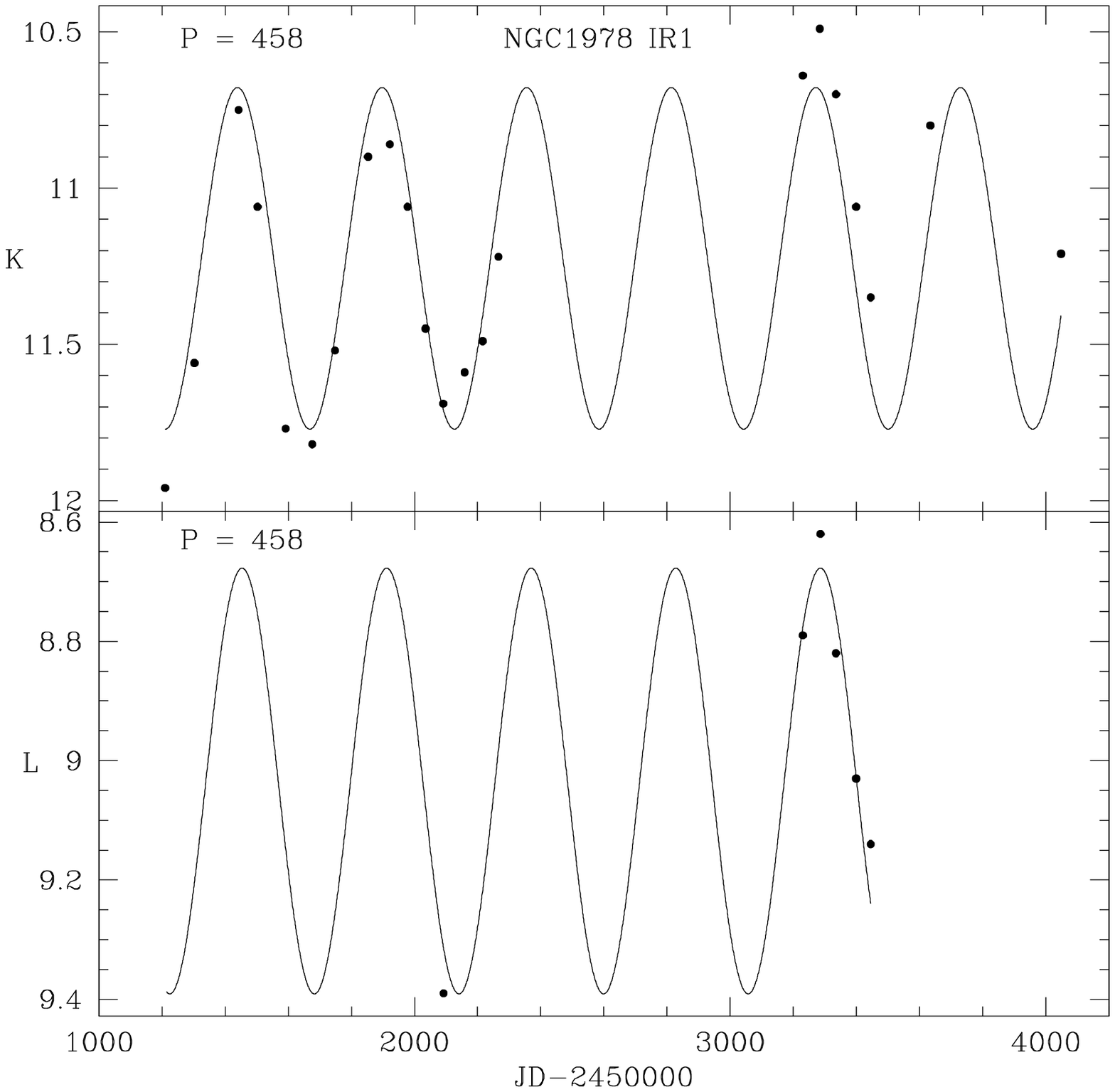}
\end{minipage}
\begin{minipage}{200pt}
\includegraphics[width=2.75in]{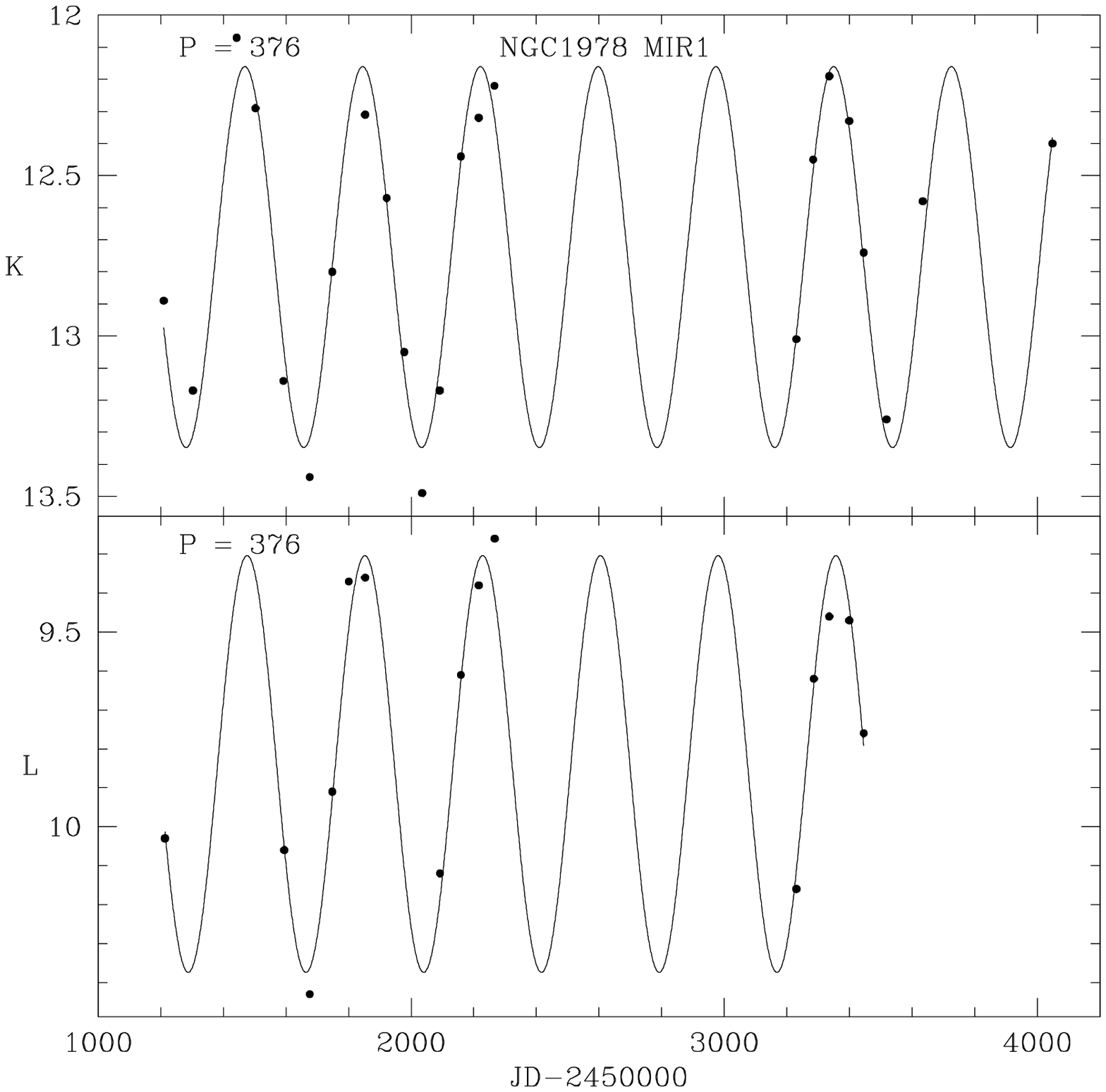}
\end{minipage}
\end{center}
\caption{$K$ and $L$ band light curve of the IR1 and MIR1 variables in NGC\,1978.
The lines are Fourier fits using the adopted period.}
\label{fig:IR1_MIR1_1978}
\end{figure*}

\begin{figure*}
\centering \includegraphics[width=150mm]{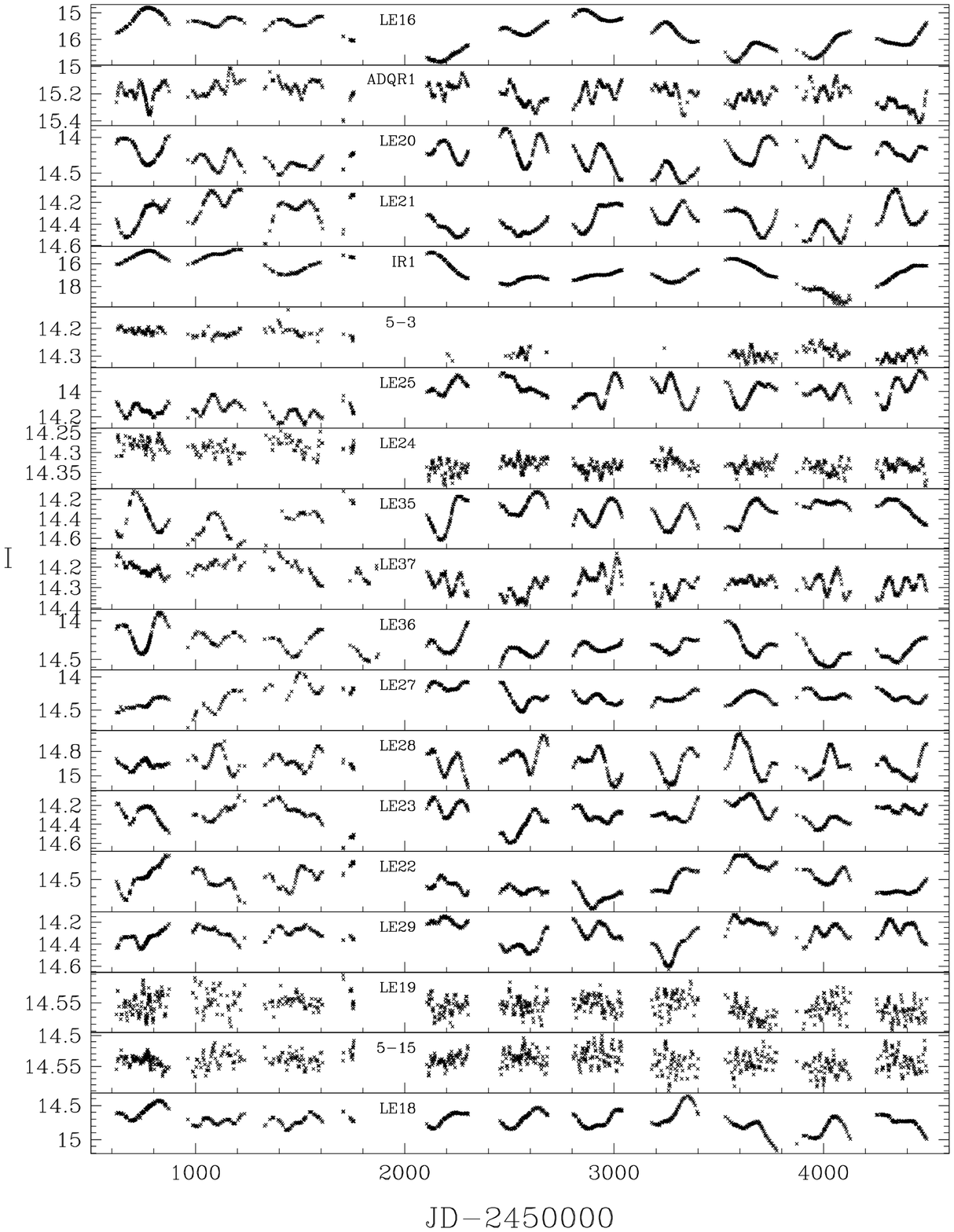}
 \caption{OGLE $I$ band light curves of the variable AGB stars in NGC\,419.}
\label{fig:n419LC}
\end{figure*}

The pulsation periods for most of the target stars in the two
clusters NGC\,1978 and NGC\,419 were determined by analysing their
MACHO and OGLE light curves. The exceptions were the
near-IR and mid-IR sources in each cluster, for which the near-IR
photometric observations ($K$ and $L$ bands) were used to determine the
pulsation period.

\begin{figure*}
\begin{center}
\begin{minipage}{200pt}
\includegraphics[width=2.75in]{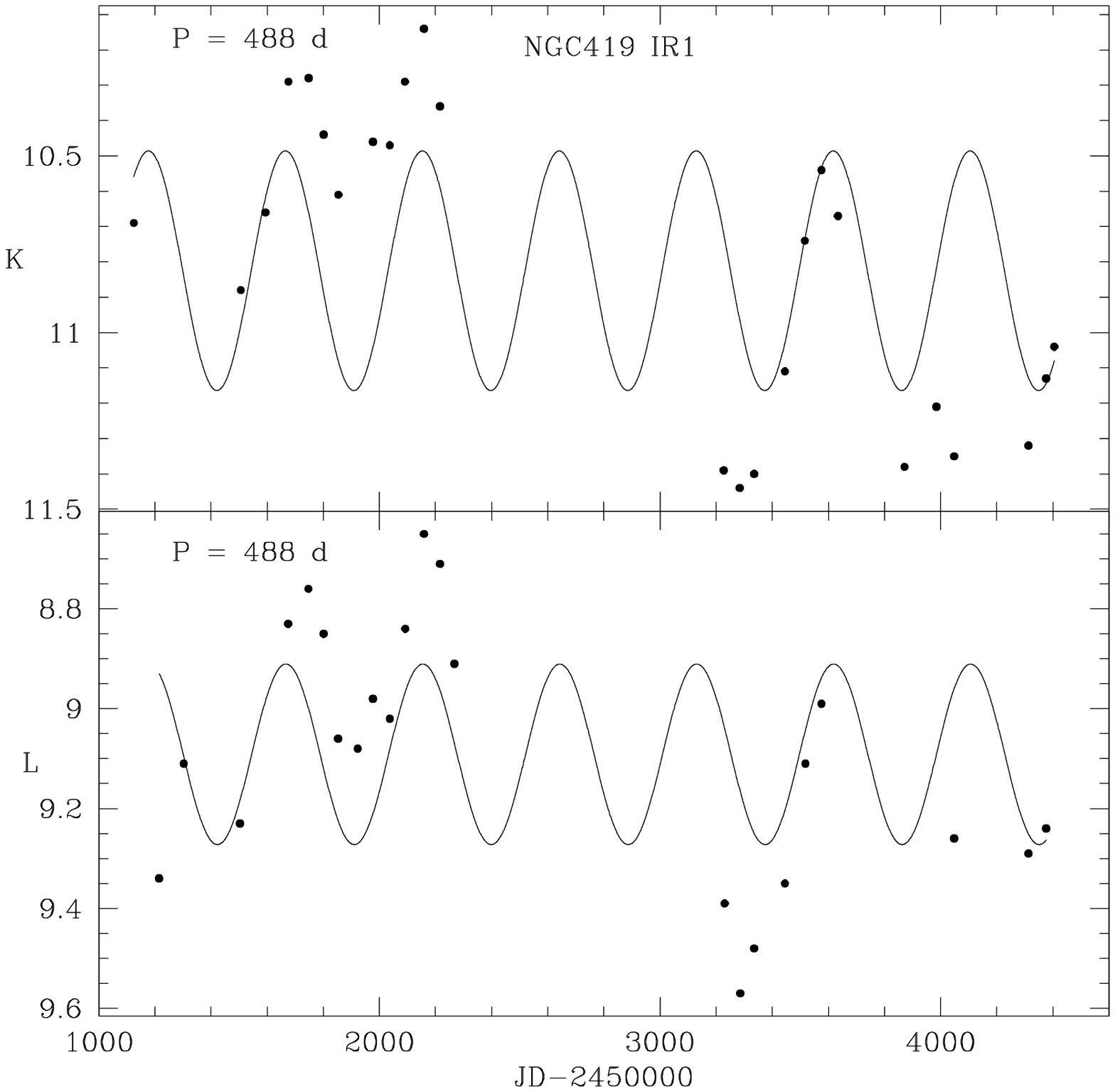}
\end{minipage}
\begin{minipage}{200pt}
\includegraphics[width=2.75in]{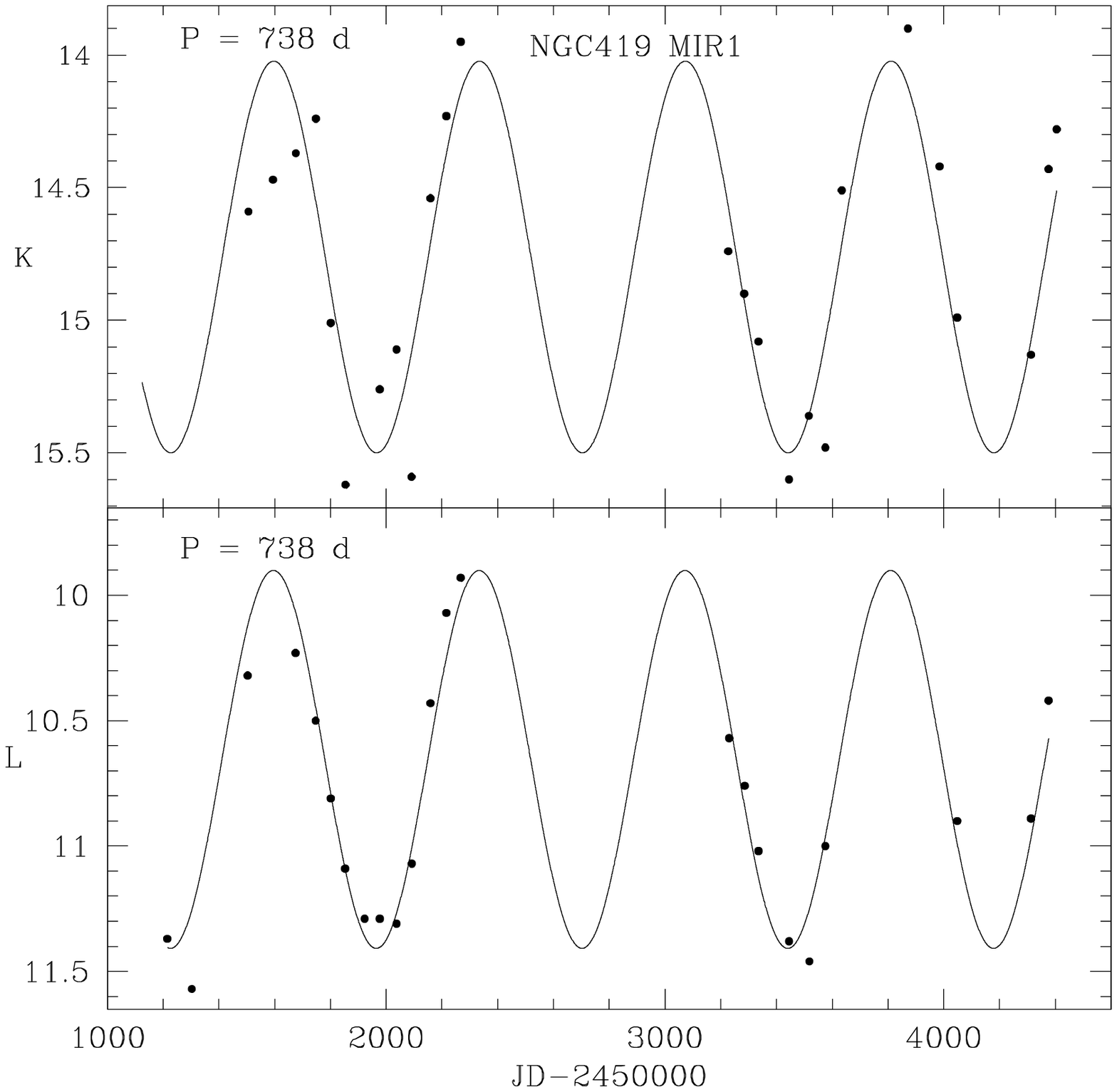}
\end{minipage}
\end{center}
\caption{$K$ and $L$ band light curve of the IR1 and MIR1 variables in NGC\,419.
The lines are Fourier fits using the adopted period.}
\label{fig:IR1_MIR1_419}
\end{figure*}

A glance at the light curves in Figures~\ref{fig:n1978LC} and \ref{fig:n419LC} shows that all the
stars show semi-regular light variations with a characteristic
period, and most of the stars are multi-periodic.  Most of the
variables have small amplitudes ($<$0.5 magnitudes). The mid-IR and near-IR
sources in each cluster have fairly large amplitudes as well as long
periods (the mid-IR sources have only the $K$ and $L$ light curves shown in
Figures~\ref{fig:IR1_MIR1_1978} and \ref{fig:IR1_MIR1_419}).   

In order to determine the periods, the light curve for each of the
candidates was first visually inspected to arrive at an approximate
period or periods.  Once a rough estimate of the period was made, two
more quantitative period estimates were made for each star.  First,
the light curve was analysed using the Phase Dispersion Minimization
(PDM) task in IRAF and a period or periods were determined.  Then the
Period04\footnote{www.univie.ac.at/tops/Period04/} software, an extended version of Period98 by Sperl (1998),
was used for period determination.  Period04 makes multi-period
Fourier fits to the light curve - a maximum of three periods was
allowed in our fitting.  In many cases, it is far from clear that the fit
periods are real. The stellar oscillations in our stars have short coherence times and the
periods often seem to vary, possibly due to changes in the envelope
structure associated with convection in the stellar envelope.  We have only retained
periods that appear in both the Period04 and PDM analyses, as well as
being stable and apparent to the eye.

For the large amplitude MIR1 and IR1 variables in each cluster, the
periods were determined from Fourier fits to the $K$ and $L$ light
curves (see Figures~\ref{fig:IR1_MIR1_1978} and
\ref{fig:IR1_MIR1_419}).  \textbf{Nishida et al. (2000) also derived
the periods of the IR1 variables in the two clusters from K
light-curves.  However, owing to the relatively short interval of
their data ($\sim$500 days in NGC\,1978 and $\sim$800 days in
NGC\,419), the periods measured by Nishida et al. (2000) (P =
491 days for NGC\,1978 IR1 and P = 526 days for NGC\,419 IR1) are
rather approximate.}

\textbf{We also note that Nishida et al. found a mean K magnitude for NGC\,1978 IR1 
of 10.51, 0.62 magnitudes brighter than our mean K (11.13), while their mean K 
magnitude for NGC\,419 IR1 is 11.10 which is 0.34 magnitudes fainter than 
our value (10.76).  The variability of the mean K magnitude is also apparent 
in our long (~$\sim$3300 day) light curves.  These long term trends appear to 
be cyclic (for example, see the light curves of LE16, LE20 and IR1 in 
Figure~\ref{fig:n419LC}) and they are common features in the light curves of dust 
enshrouded AGB stars (e.g. Whitelock et al. 2003; Groenewegen et al. 2007).  
It is unclear whether these long cyclic variations are the same as those 
seen in less extreme AGB stars which show long secondary periods on 
sequence-D in the period-luminosity diagrams of variable red 
giants (e.g. Wood et al. 1999).}

Table 3 and Table 4 give the periods determined for each of the
variables.  Periods that are considered uncertain are followed by a
colon.  A few stars, LE4 in NGC\,1978 and LE16 and LE29 in
NGC\,419, show a dominant long secondary period, a well known feature
of AGB variables.  For LE23 in NGC\,419, variability was detected but
no period could be determined.  We note that Tanab\'e et al. (2004)
determined periods for most of the NGC\,419 variables from much
shorter segments of OGLE light curves (up to JD = 2452000). For stars 
that occur in both our study and that of Tanab\'e et al. (2004), the 
periods we determined were also found by  Tanab\'e et al. (2004) in only 
9 out of 21 cases (with period agreement to 10\% or better).

\section[]{Bolometric Magnitudes and Luminosities}
\label{sec:BML} 

\begin{figure}
\centering
 \includegraphics[width=80mm]{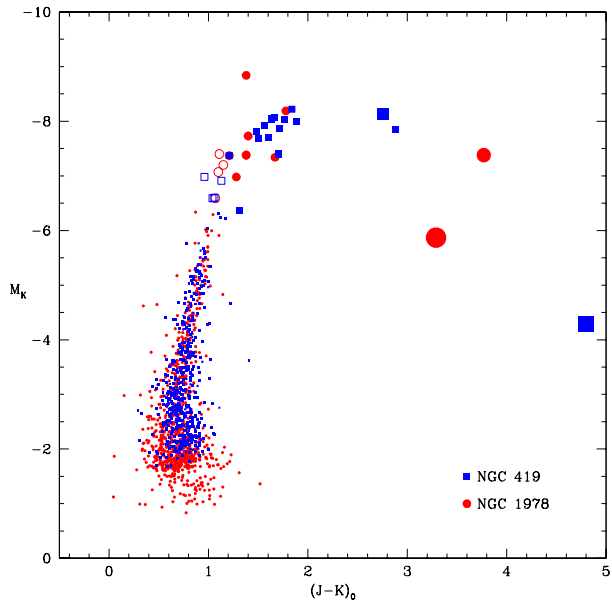} 
\caption{The $M_{\rmn{K}}$,($J$-$K$)$_{\rmn{0}}$ diagram for red giant stars 
in NGC\,1978 and NGC\,419. The variable stars are shown as larger symbols, 
circles for NGC\,1978 and squares for NGC\,419. The open symbols represent the variable M stars 
while the filled symbols depict the variable C stars. The intermediate size symbols 
represent the IR1 variables and the two largest symbols correspond to 
the MIR1 variables.}
 \label{fig:mkjmk}
\end{figure}

\begin{table*}
 \begin{minipage}{170mm}
\caption{Intrinsic properties of the AGB variables in NGC\,1978.  
Periods are listed in days. Multiple periods are listed on consecutive lines. 
Uncertain periods are marked with a colon. The last column gives 
the source of the mid-IR observations. }
   \centering
\renewcommand{\thefootnote}
  \centering
  \begin{tabular}{@{}lccccccccc}
   \hline
   Ident & $P$ & $J_{\rmn{0}}$ & $K_{\rmn{0}}$ & ($J$-$K$)$_{\rmn{0}}$ & $L$/\Lsolar & 
$M_{\rmn{bol}}$\footnote{($M_{\rmn{bol}}$)$_{\rmn{\odot}}$ = 4.72} & $\log T_{\rmn{eff}}$ & C/O\footnote{From Lederer et al. (2009)} 
& MIR\footnote{Source of MIR data: g for Groenewegen et al. (2007) and s for the SAGE survey, Meixner et al. (2006)} \\ 
  \hline
LE3  & 474 & 11.01 &  9.67 & 1.34 & 11111 & -5.394 & 3.503 & $>$1.50       & g \\
LE6  & 136 & 12.14 & 10.78 & 1.36 &  6688 & -4.843 & 3.499 & 1.30$\pm$0.10 & s \\
LE4  &  77 & 12.42 & 11.31 & 1.11 &  4613 & -4.440 & 3.551 & 0.18$\pm$0.03 & s \\
     & 654\footnote[3]{Long Secondary Period }\\
B    & 100 & 12.31 & 11.14 & 1.17 &  5019 & -4.532 & 3.538 & 1.35$\pm$0.10 & s \\
LE5  &  47 & 12.50 & 11.44 & 1.06 &  4024 & -4.292 & 3.562 & 0.18$\pm$0.05 & s \\
IR1  & 458 & 14.86 & 11.13 &  --- &  6988 & -4.891 &  ---   &               & g \\
A    &  48 & 12.18 & 11.11 & 1.07 &  5870 & -4.702 & 3.559 & 0.23$\pm$0.05 & s \\
LE11 & 110 & 12.77 & 11.53 & 1.24 &  3576 & -4.164 & 3.523 &               & s \\
LE1  & 196 & 12.06 & 10.32 & 1.74 &  8153 & -5.058 & 3.431 & $>$1.50       & s \\
LE8  &  32 & 12.95 & 11.92 & 1.03 &  2907 & -3.939 & 3.568 &               & s \\
LE7  & 178 & 12.47 & 11.13 & 1.34 &  4754 & -4.473 & 3.503 &               & s \\
     & 108: \\
LE2  & 121 & 12.80 & 11.17 & 1.63 &  4035 & -4.295 & 3.450 &               & s \\
MIR1 & 376 & 15.89 & 12.64 &  --- &  5488 & -4.628 &  --- &               & g \\
 \hline
 \end{tabular} 
 \end{minipage}
\end{table*}

Absolute bolometric magnitudes were obtained by assuming an LMC
distance modulus of 18.54 and a SMC distance modulus of 18.93 (Keller
\& Wood 2006).  Keller \& Wood also derived a mean reddening $E(B$-$V)
= 0.08$ for the LMC and $E(B$-$V) = 0.12$ for the SMC and we use these
values. For all the AGB variables the photometry was corrected
for foreground extinction using the extinction law of Cardelli et
al. (1989). The bolometric luminosity was then obtained by integrating
under the spectral energy distribution defined by the photometry in
Table 1 and Table 2 together with SST mid-IR data from either IRAC or
the IRS as noted in Tables 3 and 4.  The IR1 and MIR1 stars in each
cluster are of large amplitude so a correction needs to be made for
the fact that they were observed only once in the mid-IR.  For these
stars, $M_{\rm{bol}}$ was computed at the time of the SST observation
using the mid-IR observations and the near-simultaneous $JHKL$
observations listed in Groenewegen et al. (2007).  It was then assumed
that the $M_{\rm{bol}}$ and the $L$ magnitude have the same amplitude
of variation and the computed values of $M_{\rm{bol}}$ at the time of
the SST observation was then corrected to the flux-mean value by using the
difference between $L$ at the time of the SST observation and the mean
$L$.  This correction was particularly large for NGC\,419 MIR1 which
was observed near minimum and needed a correction of 0.72 magnitudes.
The reddening corrected photometry along with the luminosities and the
bolometric magnitudes for the target
stars in NGC\,1978 and NGC\,419 are listed in Table 3 and Table 4,
respectively.

Effective temperatures $T_{\rm eff}$ were also estimated for the variable stars.
For the O-rich stars in each cluster ($J$-$K$)$_{\rm{0}}$ was converted to 
$T_{\rm{eff}}$ using the transforms in Houdashelt et al. (2000a,b).
For the C-rich stars, ($J$-$K$)$_{\rm{0}}$
was converted to $T_{\rm{eff}}$, using the
($J$-$K$)$_{\rm{0}}$,$T_{\rm{eff}}$ relation from Bessell et al. (1983).
These $T_{\rm eff}$ values are given in Tables 3 and 4.

Figure~\ref{fig:mkjmk} shows the $M_{\rm K}$, ($J$-$K$)$_{\rm 0}$
diagram for variable and non-variable red giants in NGC\,1978 and
NGC\,419.  All the variables lie at the tip of the red giant branch.
As expected, the C stars are redder and more luminous than the M stars
in each of the two clusters. For both clusters, a distinct M to C
transition can be observed at a ($J$-$K$)$_{\rm 0} \sim 1.3$.  The
very red, high mass loss rate stars have fainter $K$ magnitudes
because most of their energy is emitted in the mid-IR, at wavelengths
longer than the $K$ band.

\begin{table*}
 \begin{minipage}{170mm}
   \caption{Intrinsic properties of the AGB variables in NGC\,419.  
Periods are listed in days. Multiple periods are listed on consecutive lines. 
Uncertain periods are marked with a colon. The last column gives the source of 
the mid-IR observations.}
\centering
\renewcommand{\thefootnote}{\thempfootnote}
\centering
  \begin{tabular}{@{}lcccccccc}
  \hline
   Ident & $P$ & $J_{\rmn{0}}$ & $K_{\rmn{0}}$ & ($J$-$K$)$_{\rmn{0}}$ & $L$/\Lsolar & 
$M_{\rmn{bol}}$\footnote{($M_{\rmn{bol}}$)$_{\rmn{\odot}}$ = 4.72} & $\log T_{\rmn{eff}}$ & 
MIR\footnote{Source of MIR data: g for Groenewegen et al. (2007) and s for 
the S$^3$MC survey, Bolatto et al. (2007)} \\ 
  \hline
LE16  & 207 & 13.85 & 11.04 & --- &  5172 & -4.564 & --- & g \\
      & 1803\footnote{Long Secondary Period }\\
      & 420: \\
ADQR1 &  62: & 13.76 & 12.52 & 1.24 &  2155 & -3.613 & 3.523 & s \\
LE20  & 151 & 12.45 & 10.96 & 1.49 &  7327 & -4.942 & 3.475 & s \\
LE21  & 163 & 12.56 & 10.86 & 1.70 &  7235 & -4.929 & 3.438 & s \\ 
      & 317 \\
IR1   & 488 & 13.45 & 10.76 &  --- &  8841 & -5.146 &  ---  & g \\
5-3   &  40 & 12.80 & 11.91 & 0.89 &  4634 & -4.445 & 3.601 & s \\
LE25  & 133 & 12.48 & 11.07 & 1.41 &  6978 & -4.889 & 3.490 & s \\
      & 238: \\
LE24  &  38 & 13.04 & 11.98 & 1.06 &  4124 & -4.318 & 3.562 & s \\
LE35  & 175 & 12.41 & 10.82 & 1.59 &  5494 & -4.630 & 3.457 & g \\
LE37  &  85 & 12.66 & 11.52 & 1.14 &  5435 & -4.618 & 3.544 & s \\
LE36  & 185 & 12.44 & 10.67 & 1.77 &  8682 & -5.127 & 3.426 & s \\
LE27  & 166: & 12.41 & 10.85 & 1.56 &  4713 & -4.463 & 3.462 & g \\
LE28  & 136: & 13.12 & 11.49 & 1.63 &  4504 & -4.414 & 3.450 & s \\
LE23  &  -- & 12.72 & 11.19 & 1.53 &  6188 & -4.759 & 3.467 & s \\
LE22  & 160 & 12.67 & 11.02 & 1.65 &  6223 & -4.765 & 3.446 & s \\
LE29  & 126 & 12.63 & 11.20 & 1.43 &  6369 & -4.790 & 3.486 & s \\
      & 658\footnotemark[\value{mpfootnote}]\\
LE19  &  31 & 13.27 & 12.30 & 0.97 &  3290 & -4.073 & 3.582 & s \\
5-15  &  31 & 13.28 & 12.29 & 0.99 &  3313 & -4.080 & 3.578 & s \\ 
LE18  & 181 & 12.71 & 10.89 & 1.82 &  4993 & -4.526 & 3.418 & g \\
MIR1  & 738 &  0.00 & 14.60 &  --- & 10481 & -5.331 &  ---  & g \\
 \hline
 \end{tabular}
  \end{minipage}
\end{table*}

\section[]{Linear pulsation models}

\begin{figure}
\centering
 \includegraphics[width=80mm]{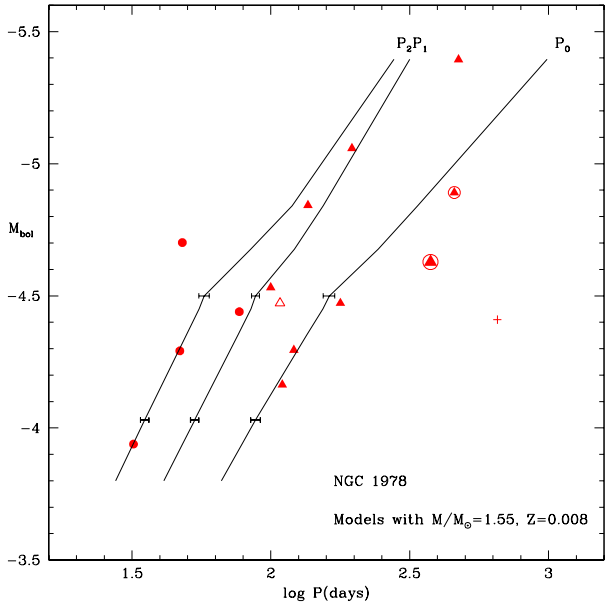}
 \caption{The $M_{\rmn{bol}}$,$\log P$ diagram for red giant stars 
in NGC\,1978. The filled circles denote the M stars and 
the closed triangles denote the C stars.  The circled triangles
are the sources IR1 and MIR1, with the latter having a larger symbol.  The open symbol corresponds 
to the uncertain period in LE7 (as indicated in Table 3). The plus symbol represents the 
long secondary period in LE4.  The lines are theoretical period-luminosity
relations for small amplitude stars pulsating in the fundamental mode and
the first and second overtone.  The error bars show the effect of
changing the fitted mass by 0.1\,M$_{\odot}$.}
 \label{fig:mbolp_n1978}
\end{figure}

\begin{figure}
\centering
\includegraphics[width=80mm]{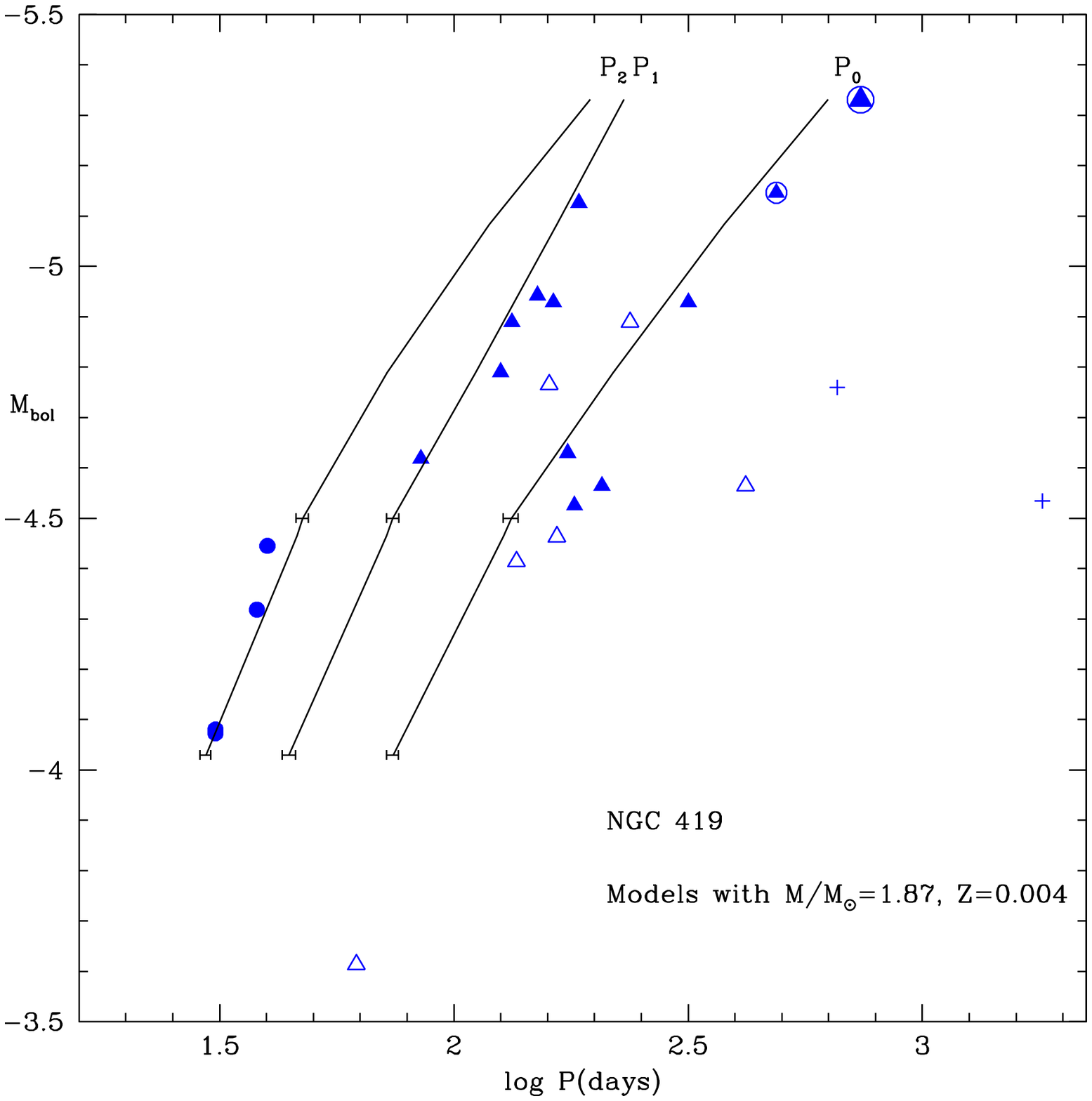}
 \caption{The $M_{\rmn{bol}}$,$\log P$ diagram for red giant stars 
in NGC\,419. The filled circles denote the M stars and 
the closed triangles denote the C stars. The circled triangles
are the sources IR1 and MIR1, with the latter having a larger symbol.  The open symbols correspond 
to stars with uncertain periods (as indicated in Table 4). The plus symbols represents the 
long secondary periods in LE16 and LE29.  The lines are theoretical period-luminosity
relations for small amplitude stars pulsating in the fundamental mode and
the first and second overtone.  The error bars show the effect of
changing the fitted mass by 0.1\,M$_{\odot}$.}
\label{fig:mbolp_n419}
\end{figure}

For the small amplitude variables in NGC\,1978 and NGC\,419 we
constructed static stellar models and linear non-adiabatic pulsation
models which are based on the pulsation codes described in Fox \& Wood
(1982) and updated by Keller \& Wood (2006).  The core mass
$M_{\rm{c}}$ was obtained from the $L$,$M_{\rm{c}}$ relation of Wood \&
Zarro (1981).  In a further update on these codes, we have used the
Rosseland mean opacities computed using AESOPUS (Marigo \& Aringer
2009) at temperatures below $\log T = 3.75$.  This allowed us to
construct models with different C/O ratios.  We used a (solar-scaled)
metallicity $Z = 0.008$ for the LMC cluster NGC\,1978 and $Z = 0.004$
for the SMC cluster NGC\,419.

To get the parameters for the static stellar models in each of the two
clusters, we first constructed a fiducial giant branch for the
non-variable AGB stars in the $M_{\rm{bol}}$,$T_{\rm{eff}}$ diagram.
Our starting point was the $M_{\rm{K}}$,($J$-$K$)$_{\rm 0}$ diagram
shown in Figure~\ref{fig:mkjmk}.  In each cluster, the non-variable
stars are all O-rich and $M_{\rm{K}}$ and ($J$-$K$)$_{\rm{0}}$ were
converted to $\log$($L$/\Lsolar)\ and $T_{\rm{eff}}$ using the
transforms in Houdashelt et al. (2000a,b).  By adjusting the mixing
length in the convection theory, the static models were made to match
the observed non-variable AGB in each cluster i.e. the observed
$T_{\rm{eff}}$ of the O-rich stars on the AGB was reproduced. The
mixing length was then kept constant for higher luminosity models,
both O- and C-rich models, and linear and nonlinear.
 
A significant uncertainty in our models is the C/O ratio to be used.
We started with a C/O $\sim$ 0.30 low on the AGB for the two clusters
(the original solar-scaled C and N abundances were modified to
allow for CN cycle conversion of C to N to achieve this ratio).
This ratio is close to that found in standard stellar evolution
calculations for low mass stars (e.g. Girardi et al. 2000).  However,
we note that this ratio is higher than the typical value C/O $\sim$
0.18 found by Lederer et al. (2009) for O-rich AGB stars in NGC\,1978 (see Table 3).
Fortunately, O-rich models are insensitive to this uncertainly in the
C/O ratio.  In both clusters, the M to C star transition takes place
near $M_{\rm{bol}}$ = -4.5.  This can be seen in Figures~\ref{fig:mbolp_n1978} 
and \ref{fig:mbolp_n419} respectively, where we plot bolometric luminosities 
of AGB variables versus $\log P$.  We therefore assume
C/O = 1.00 at this luminosity.  We made a synthetic AGB calculation
similar to that in Marigo et al. (1999) to estimate the C/O ratio
increase with $M_{\rm{bol}}$ up to 1.0 at $M_{\rm{bol}}$ = -4.5.  The
change in C/O with $M_{\rm{bol}}$ in our models is given in Tables 5
and 6.

For the carbon stars brighter than $M_{\rm{bol}}$ = -4.5, estimating
the C/O ratio is difficult.  In NGC\,1978, Lederer et al. (2009) found
C/O $\sim$ 1.3 for two C stars and they also found two other C stars
with lower limits C/O $>$ 1.5.  In a broad sense, C/O ratios of
planetary nebulae in the LMC put upper limits on the C/O ratio that
would be appropriate for the most luminous C stars in the final heavy
mass loss stage immediately preceding the planetary nebula (PN) stage
(like the mid-IR sources in NGC\,1978 and NGC\,419).  In the LMC,
Stanghellini et al. (2005) found that a C/O of 2 is typical for
non-bipolar PN in the LMC.  We take this as the upper limit to C/O for
our models for NGC\,1978.  In NGC\,419, we have no observational
constraints on C/O in the cluster C stars.  We adopt the same upper
limit C/O = 2 in NGC\,419 as in NGC\,1978 although we note that in the
SMC, Stanghellini et al. (2009) found that a C/O of 4 is typical for
non-bipolar SMC PN.  We assume that from $M_{\rm{bol}}$ = -4.5, where
C/O = 1, the C/O ratio increases to 2.0 at the luminosity of the most
luminous star in each cluster.  The luminosities and C/O ratios of the
linear models are given in Tables 5 and 6.

In order to derive the pulsation mass, the mass used in the linear
pulsation models was adjusted so that the models reproduced the period
of the lowest luminosity ($M_{\rm bol} \sim -4$) O-rich star in each
cluster.  These stars should have undergone the least mass loss, and
their observed periods are well defined.  Changes to the mass of the
model require a refinement in the value of the mixing-length parameter
so that simultaneous fitting of both $T_{\rm{eff}}$ for the fiducial
giant branch and the period of the lowest luminosity pulsator is an
iterative process.  Once the mass was determined, it was kept constant
for the more luminous models (the mixing-length was also kept
constant).  Under this assumption, the periods of the three lowest
order modes were computed and their values are given in Tables 5 and 6
while the lines in Figures~\ref{fig:mbolp_n1978} and
\ref{fig:mbolp_n419} show the variation of the periods with
luminosity.  To get an estimate of variation of the model periods with
the fitted mass, masses $\rm$0.1\,M$_{\odot}$ different from the
optimum value were also used with the above procedure (i.e. the
mixing-length was adjusted to give the fiducial $T_{\rm{eff}}$) and
their model periods were computed.  The resulting change in the
periods are shown as error bars in Figures~\ref{fig:mbolp_n1978} and
\ref{fig:mbolp_n419}.

In both clusters, the lowest luminosity O-rich stars pulsate in the second
overtone mode.  The C stars pulsate in a mixture of fundamental and
first overtone modes.  This agrees with the finding of
previous studies that stars start pulsating on the AGB in higher overtones
and they move to lower order modes, ultimately the fundamental mode,
as luminosity increases (Lebzelter \& Wood 2005;\,2007).

The best pulsation mass low on the AGB for NGC\,1978 is
1.55\,M$_{\odot}$ in good agreement with the mass of 1.44 to 1.53
\,M$_{\odot}$ for early AGB stars predicted by isochrones (see
Section\,1).  Similarly, the best pulsation mass low on the AGB for
NGC\,419 is 1.87\,M$_{\odot}$ in very good agreement with the mass of
1.64 to 1.94 \,M$_{\odot}$ predicted by isochrones.

\begin{table*}
 \begin{minipage}{170mm}
   \caption{Linear pulsation models - NGC\,1978}
\centering
\renewcommand{\thefootnote}{\thempfootnote}
\centering
  \begin{tabular}{@{}lccccccccc}
  \hline
   $M_{\rm bol}$ & $M$/\Msolar & $M_{\rm{c}}$/\Msolar & $\ell$/H$_{\rm{p}}$ & $\log T_{\rm{eff}}$ & C/O & $P_{\rm{0}}$ & $P_{\rm{1}}$ & $P_{\rm{2}}$ & $P_{\rm{3}}$\\ 
  \hline
-3.80 & 1.55 & 0.538 & 2.255 & 3.5731 & 0.33 &  66.3 &  41.2 &  27.6 &  21.2 \\
-4.35 & 1.55 & 0.567 & 2.255 & 3.5455 & 0.50 & 134.8 &  76.2 &  49.7 &  42.6 \\
-4.45 & 1.55 & 0.574 & 2.255 & 3.5407 & 0.80 & 154.1 &  84.8 &  55.2 &  48.3 \\
-4.50 & 1.55 & 0.577 & 2.255 & 3.5408 & 0.10 & 162.3 &  88.1 &  57.5 &  50.9 \\
-4.68 & 1.55 & 0.591 & 2.255 & 3.5115 & 1.20 & 243.3 & 121.6 &  84.6 &  74.3 \\
-4.84 & 1.55 & 0.608 & 2.255 & 3.4913 & 1.50 & 337.5 & 154.5 & 119.4 &  96.2 \\
-5.39 & 1.55 & 0.683 & 2.255 & 3.4328 & 2.00 & 987.5 & 316.5 & 277.5 & 221.8 \\
\hline
 \end{tabular}
  \end{minipage}
\end{table*}

\begin{table*}
 \begin{minipage}{170mm}
   \caption{Linear pulsation models - NGC\,419}
\centering
\renewcommand{\thefootnote}{\thempfootnote}
\centering
  \begin{tabular}{@{}lccccccccc}
  \hline
   $M_{\rm bol}$ & $M$/\Msolar & $M_{\rm{c}}$/\Msolar & $\ell$/H$_{\rm{p}}$ & $\log T_{\rm{eff}}$ & C/O & $P_{\rm{0}}$ & $P_{\rm{1}}$ & $P_{\rm{2}}$ & $P_{\rm{3}}$\\ 
  \hline
-4.03 & 1.87 & 0.548 & 1.845 & 3.5717 & 0.31 &  74.2 &  44.5 &  29.6 &  22.6 \\
-4.43 & 1.87 & 0.572 & 1.845 & 3.5529 & 0.50 & 122.5 &  69.2 &  44.7 &  36.0 \\
-4.46 & 1.87 & 0.575 & 1.845 & 3.5517 & 0.80 & 127.9 &  71.8 &  46.3 &  37.5 \\
-4.50 & 1.87 & 0.577 & 1.845 & 3.5513 & 1.00 & 132.7 &  74.0 &  47.6 &  38.5 \\
-4.79 & 1.87 & 0.602 & 1.845 & 3.5238 & 1.20 & 218.7 & 111.3 &  72.0 &  64.6 \\
-5.08 & 1.87 & 0.636 & 1.845 & 3.4944 & 1.50 & 378.0 & 166.1 & 119.2 & 100.6 \\
-5.33 & 1.87 & 0.671 & 1.845 & 3.4678 & 2.00 & 629.4 & 230.8 & 195.4 & 150.5 \\
\hline
 \end{tabular}
  \end{minipage}
\end{table*}

\section[]{Nonlinear pulsation models and mass loss}

\begin{figure}
\resizebox{\hsize}{!}{\includegraphics[clip=true]{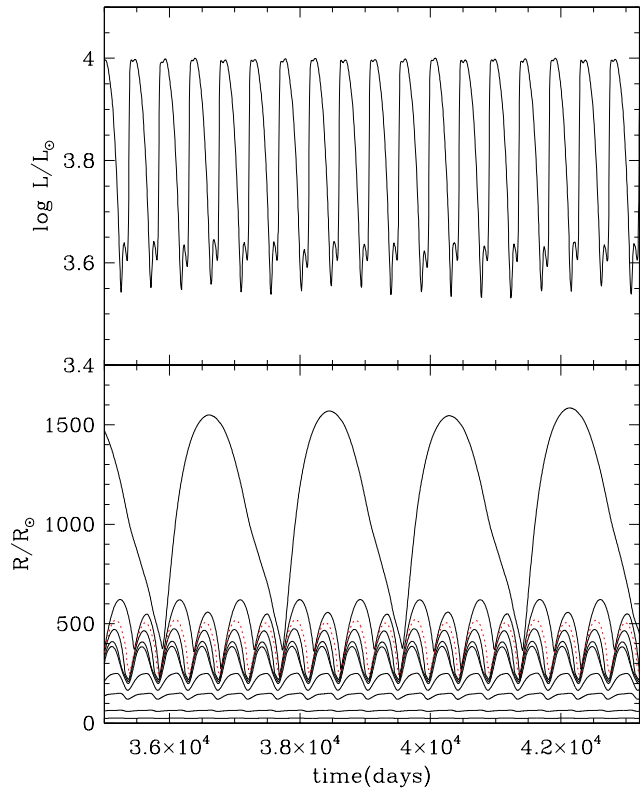}}
 \caption{Nonlinear pulsation model for the IR source in NGC\,1978.  The top
panel shows the time variation of the surface luminosity.  The black lines
in the bottom panel show the radius variation of selected mass points in
the envelope while the red dotted line shows the position of the point where
the Rosseland mean optical depth is 2/3.  The model
parameters are given at the top of the plot.}
\label{fig:ngc1978_ir1_model}
\end{figure}

\begin{figure}
\resizebox{\hsize}{!}{\includegraphics[clip=true]{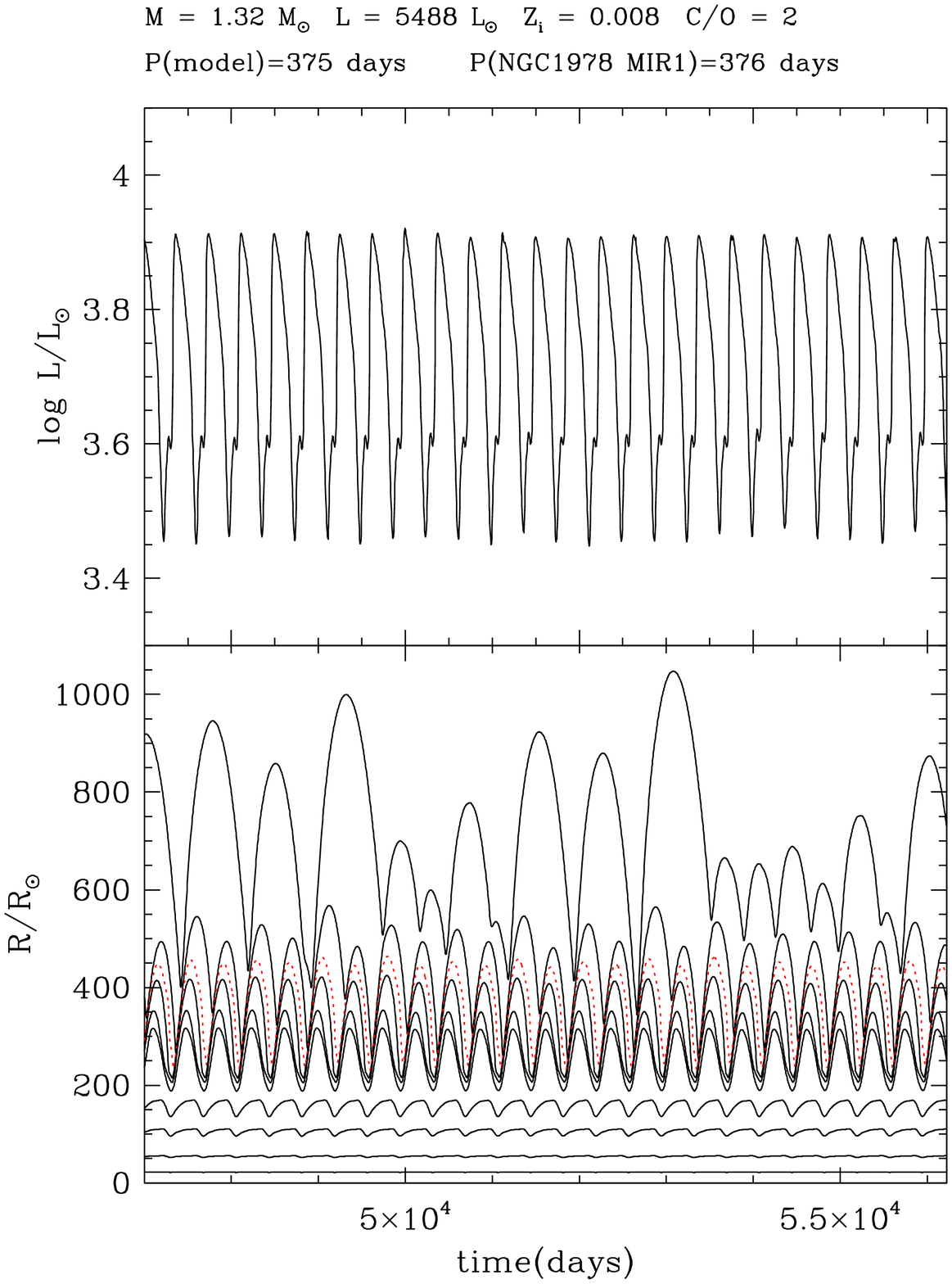}}
 \caption{The same as Figure~\ref{fig:ngc1978_ir1_model} but for the MIR source in NGC\,1978.}
\label{fig:ngc1978_mir1_model}
\end{figure}

\begin{figure}
\resizebox{\hsize}{!}{\includegraphics[clip=true]{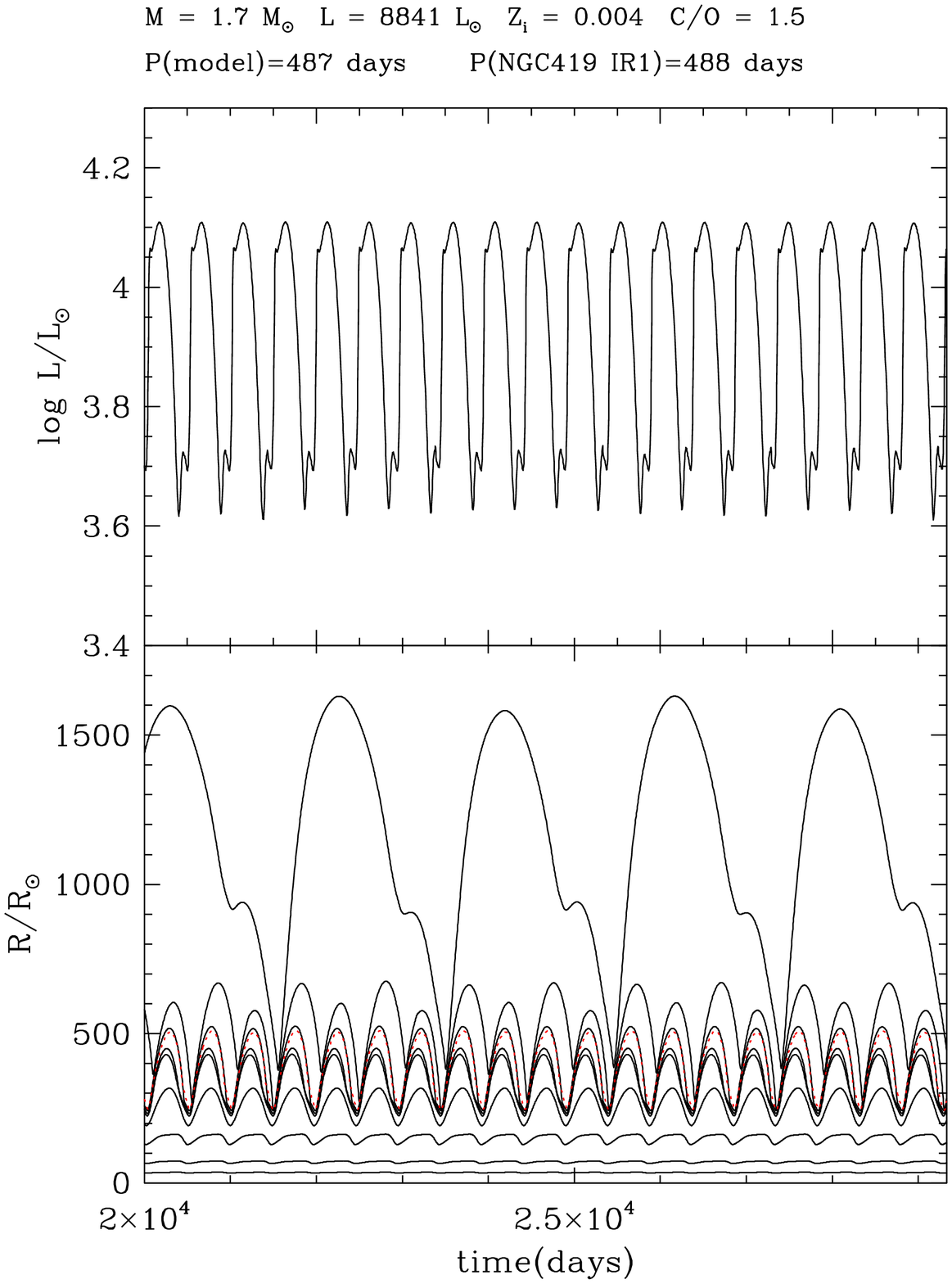}}
 \caption{The same as Figure~\ref{fig:ngc1978_ir1_model} but for the IR source in NGC\,419.}
\label{fig:ngc419_ir1_model}
\end{figure}

\begin{figure}
\resizebox{\hsize}{!}{\includegraphics[clip=true]{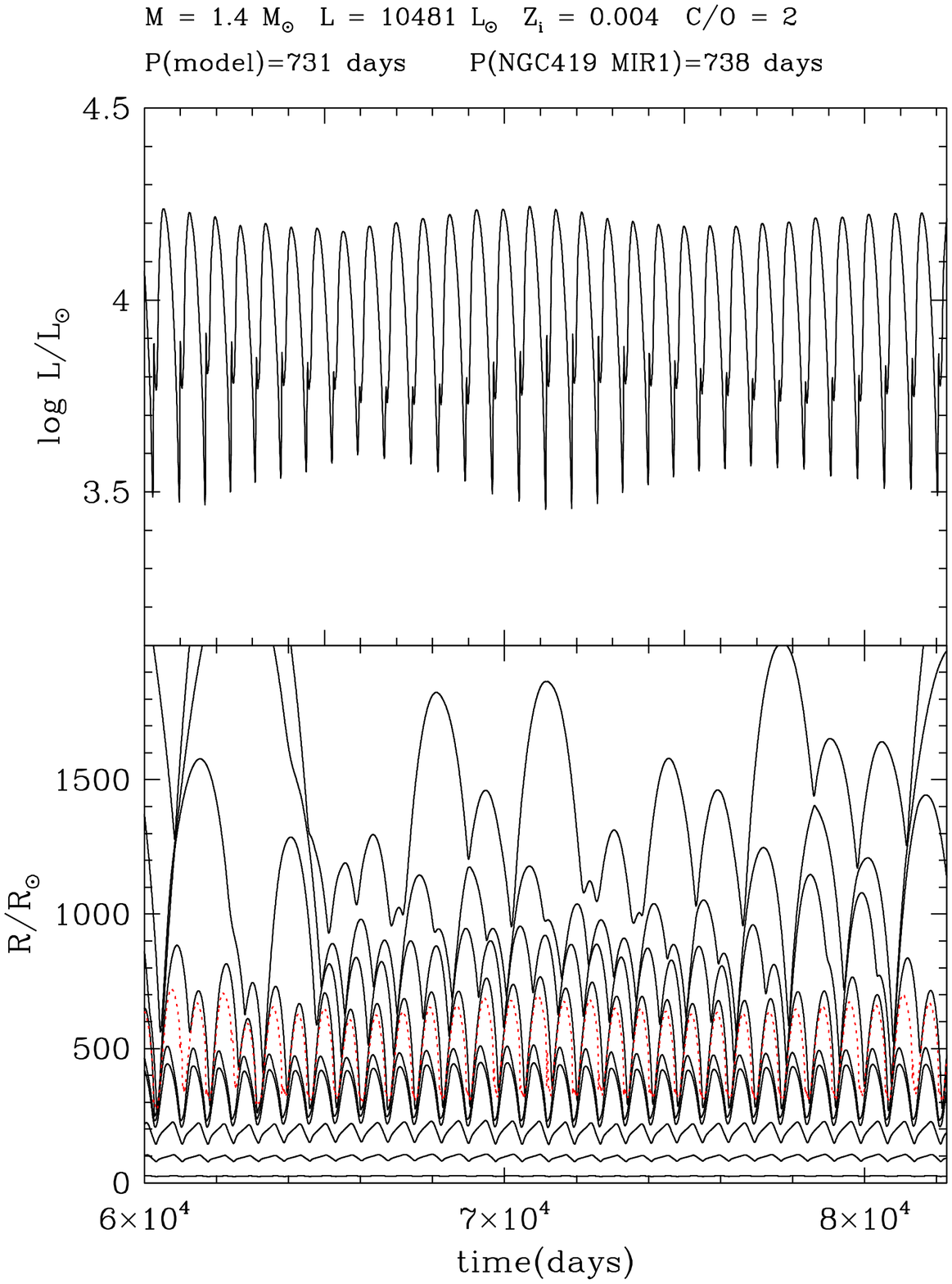}}
 \caption{The same as Figure~\ref{fig:ngc1978_ir1_model} but for the mid-IR source in NGC\,419.}
\label{fig:ngc419_mir1_model}
\end{figure}

The near-IR and mid-IR sources in the two clusters are very red and
clearly have large mass loss rates.  This is confirmed by the mass
loss rates derived for these stars by Groenewegen et al. (2007).  They
should therefore have lost considerable amounts of mass during their
AGB evolution.  By determining their current masses using pulsation
models, we can estimate how much mass has been lost, and we can
compare this to what we expect would have been lost according to
commonly-used mass loss formulae such as those of Vassiliadis \& Wood
(1993) and Bl\"ocker (1995).  Since these stars are large amplitude
pulsators, we need to model them with nonlinear pulsation models
because the linear and nonlinear pulsation periods for AGB stars can
be very different (e.g. Lebzelter \& Wood 2005).  We now discuss the
nonlinear pulsation models in each of the clusters.

\subsection{NGC\,1978}

As seen in Figure~\ref{fig:mbolp_n1978}, the sources IR1 and MIR1 in
NGC\,1978 are neither the most luminous nor the longest-period AGB
stars in this cluster even though they are the reddest by a long
margin (see Figure~\ref{fig:mkjmk}).  This is a strange situation as
the reddest stars with the largest mass loss rates should be the most
luminous and the most evolved stars in the cluster.  The most luminous
star in NGC\,1978 is LE3 but it is not particularly red.  One
possibility is that this object is a superposition on the sky of two
bright AGB stars but we can find no evidence for this: the star
appears single in an archive HST image of the cluster and Lederer et
al. (2009) did not notice any multiple velocity components in their
high resolution infrared spectra.  We therefore assume that LE3 is
indeed a single star.  There are, however, several pieces of evidence
that LE3 is not a normal cluster member: (1) it appears anomalously
bright in the $M_{\rmn{K}}$,($J$-$K$)$_{\rmn{0}}$ diagram
(Figure~\ref{fig:mkjmk}) compared to the locus formed by the other
stars in both NGC\,1978 and NGC\,419, (2) it is more luminous than any
star in NGC\,419 even though the low luminosity AGB stars in NGC\,419
have masses of 1.87\,M$_{\odot}$ compared to the smaller value of
1.55\,M$_{\odot}$ in NGC\,1978, and (3) the period of LE3 lies midway
between the fundamental and first overtone relations in
Figure~\ref{fig:mbolp_n1978}, suggesting that its mass is not normal
for the cluster.  We suggest that LE3 is a field star with a higher
mass than that of the AGB stars in NGC\,1978 or that it is an
evolved blue straggler belonging to the cluster (and hence of higher
mass than the other AGB stars).

If we ignore LE3, the most luminous cluster star becomes LE1 with
$M_{\rm bol} = -5.06$, still slightly more luminous than IR1 and
significantly more luminous that MIR1.  In order to explain the
luminosity and period of MIR1 relative to LE1, we have to assume that
MIR1 has recently undergone a helium shell flash and that it is
currently in the post-flash luminosity dip (0.43 magnitudes below the
interflash luminosity maximum).  \textbf{We note that AGB stars spend 
about a third of their time during the interpulse period 0.43 magnitudes
or more below the maximum H-burning luminosity (see the light
curves in Vassiliadis \& Wood 1993), so our assumption is reasonable.}  
IR1 would be slightly (0.16 magnitudes) below the interflash luminosity 
maximum.  In that case, the core masses for IR1 and MIR1 would 
be given by the luminosity-core mass relation for a star with the 
luminosity of LE1 at the tip of the AGB.  Hence, for our nonlinear 
pulsation models we assume the luminosity of LE1 when we derive 
the core masses from the luminosity-core mass relation. 
We note that stars thought to be in a post-flash luminosity dip 
have also been found in the intermediate age LMC cluster NGC\,1846 
(Lebzelter \& Wood 2007).

As with the linear models, a major uncertainity is the C/O ratio used
in the opacity tables.  We assume C/O = 1.5 for IR1 and C/O = 2.0 for
MIR1.  For the mixing-length, we used the same value as in the linear
models.

Given the luminosities for IR1 and MIR1 in Table 3, a series of static
models with different masses was created for each of them and these
models were then allowed to reach a full-amplitude limit cycle in the
nonlinear pulsation code.  A turbulent viscosity parameter
$\alpha_{\nu}$ (see Keller \& Wood 2006) of 0.2 was used for IR1 and
0.1 was used for MIR1 as these values gave limiting light curve
amplitudes similar to the observed amplitude.  We note that the large
amplitude pulsators go through a readjustment of their interior
structure while transitioning from a static model to a large amplitude
pulsator and that the nonlinear pulsation period can be quite
different from the linear period of the static model.  The models for
IR1 and MIR1 had linear periods of 427 and 333 days and nonlinear
periods of 459 and 375 days, respectively.

It was found that masses of 1.43\,M$_{\odot}$ and 1.32\,M$_{\odot}$
reproduced the observed periods (458 and 376 days) of IR1 and MIR1,
respectively.  The light curves and radius variations of the two
models are shown in Figures~\ref{fig:ngc1978_ir1_model} and
\ref{fig:ngc1978_mir1_model}.  Given that the mass of stars low on the
AGB of NGC\,1978 is close to 1.55\,M$_{\odot}$, it appears that IR1
has lost $\sim$0.12\,M$_{\odot}$ on the AGB while MIR1 has lost
$\sim$0.23\,M$_{\odot}$.

In order to compare the observed mass loss on the AGB on NGC\,1978
with the mass loss predicted by the commonly-used mass loss formulae
of Vassiliadis \& Wood (1993, hereinafter VW) and Bl\"ocker (1995), we
have made simple synthetic AGB evolutionary calculations including
mass loss.  Averaged over helium shell flashes, AGB stars evolve at a
constant rate in $M_{\rm bol}$ per unit time (e.g. Wood \& Cahn 1977)
and we use their rate
  \[\frac{dM_{\rm bol}}{dt} = 8.25\times10^{-7} {\rm mag~yr^{-1}} .\]
This rate is multiplied by a factor (1-$\lambda$) in the presence of
third dredge-up, where $\lambda$ is the fraction of the mass burnt
through during the previous shell flash cycle that is dredged up.  We
used $\lambda = 0.5$ as found for LMC stars by Marigo et al. (1999).
In order to calculate the mass loss rate according to VW, we need the
fundamental mode pulsation period.  Here, we use the fundamental mode
$M_{\rm bol}$,$\log P$ relation shown in Figure~\ref{fig:mbolp_n1978}
to obtain $P$: the relation is \[M_{\rm bol} = -1.95 - 1.15\log P .\]
In order to use the Bl\"ocker (1995) formula, we need the radius.
This was obtained from the fit \[\log T_{\rm eff} = 4.18 + 0.15 M_{\rm
  bol}\] to the AGB position in the theoretical HR-diagram as defined
by the $M_{\rm bol}$ and $T_{\rm eff}$ values in Table 3.

We assumed that each AGB star currently in the cluster arrived at
$M_{\rm bol} = -4$ with mass 1.55\,M$_{\odot}$.  The star was then
evolved in time using the above equations until the envelope was
dissipated.  It was immediately obvious that the Bl\"ocker (1995)
formula terminates the AGB at far too low a luminosity, $M_{\rm bol} =
-4.18$.  This is less than the luminosity of nearly all the variable
AGB stars in NGC\,1978.  On the other hand, the VW mass loss rate
caused the AGB to terminate at $M_{\rm bol} = -5.07$, in very good
agreement with the luminosity of the most luminous star LE1 at $M_{\rm
  bol} = -5.06$.  In the evolutionary models, the AGB mass reduced to the values
appropriate for IR1 and MIR1 at $M_{\rm bol} = -5.00$ and -5.03,
respectively.  This shows that mass loss is very concentrated near the tip
of the AGB.  Thus, each of the stars LE1, IR1 and MIR1 is close to the end
of its AGB lifetime which should end within the next
2$\times$10$^5$ years according to the evolutionary model.

\subsection{NGC\,419}

The IR1 and MIR1 sources in NGC\,419 occur at the red, high luminosity
tip of the AGB as expected (see Figures~\ref{fig:mbolp_n419} and
\ref{fig:mkjmk}).  Nonlinear pulsation models were created for these
two stars using the same procedures as for NGC\,1978.  Here, the core
mass was calculated from the luminosity-core mass relation using the
luminosity of the object itself (from Table 4).  As with NGC\,1978, we
assume C/O = 1.5 for IR1 and C/O = 2.0 for MIR1.

The light curves and radius variations of the two nonlinear pulsation
models are shown in Figures~\ref{fig:ngc419_ir1_model} and
\ref{fig:ngc419_mir1_model}.  The observed periods of IR1 and MIR1 in
NGC\,419 are 488 and 738 days, respectively.  The periods of the
nonlinear pulsation models are 487 and 731 days, respectively (the
corresponding linear periods of the static parent models were 509 and
1150 days).  Masses of 1.70\,M$_{\odot}$ and 1.40\,M$_{\odot}$ were
required to reproduced the observed periods of IR1 and MIR1. Given
that the mass of stars low on the AGB of NGC\,419 is close to
1.87\,M$_{\odot}$, it appears that IR1 has lost
$\sim$0.17\,M$_{\odot}$ on the AGB while MIR1 has lost
$\sim$0.47\,M$_{\odot}$ on the AGB.

As with NGC\,1978, we have made simple synthetic AGB calculations
including mass loss according to the formulae in Vassiliadis \& Wood
(1993) and Bl\"ocker (1995).  The methods used were the same as in
NGC\,1978 except that we used $\lambda = 0.65$ (the SMC value found by
Marigo et al. 1999) and the $M_{\rm bol}$,$\log P$ relation we used
was a fit to the observed fundamental mode relation given by \[M_{\rm
  bol} = -2.203 - 1.09\log P .\] In NGC\,419, as in NGC\,1978, the
Bl\"ocker (1995) formula terminates the AGB at far too low a
luminosity, $M_{\rm bol} = -4.25$, less than the luminosity of nearly
all the variable AGB stars.  The VW mass loss rate caused the AGB to
terminate at $M_{\rm bol} = -5.14$.  This is more luminous than all
stars in the cluster except MIR1 which has $M_{\rm bol} = -5.33$.
This suggests that the VW mass loss rate is slightly too large for the
stars in NGC\,419.  (However, we note that the mean luminosity of
NGC\,419 MIR1 may be more uncertain than that of most stars since the
luminosity estimated from the near-IR and SST observations needed a
particularly large increase of 0.72 magnitudes in order to get to the
mean value - see Section~\ref{sec:BML}).

The effect of the slightly low AGB termination luminosity on the
initial-final mass relation is very small since the change in core
mass between $M_{\rm bol} = -5.14$ and -5.33 is only
0.028\,M$_{\odot}$.  More significant is the effect on nucleosynthetic
yields.  The average time taken to increase $M_{\rm bol}$ from -5.14
to -5.33 on the AGB is $\sim$\,6.6$\times$10$^5$ years (if $\lambda =
0.65$), which corresponds to $\sim$\,10 shell flashes and associated
third dredge-ups.  The yields of carbon and s-process elements
returned to the interstellar medium will thus be underestimated when
using the VW mass loss rate.  Full AGB evolution calculations are
required to properly quantify these effects.

\section{Summary and conclusions}

Pulsation periods have been determined for most of the bright AGB
stars in the LMC cluster NGC\,1978 and the SMC cluster NGC\,419.  It
is shown that the AGB stars start pulsating at low luminosities on the
AGB in the second overtone mode of radial pulsation.  They switch to
lower order modes at higher luminosities and the carbon stars are
first overtone or fundamental mode pulsators.  

Pulsation masses have been determined for the O-rich pulsators low on
the AGB.  In NGC\,1978, the early-AGB mass is 1.55\,M$_{\odot}$ while
it is 1.87\,M$_{\odot}$ in NGC\,419.  These masses are in very good
agreement with the masses predicted by isochones which are consistent
with the ages of the clusters determined from main-sequence fitting.

Reasons are given that the brightest star in NGC\,1978, LE3, is not a
cluster member, or perhaps it is an evolved blue straggler.  Neither
of the two very dusty AGB stars in NGC\,1978, IR1 and MIR1, is the
brightest star on the AGB.  It is suggested that MIR1 is in
post-helium shell flash luminosity dip and that it is about 0.43
magnitudes below the interflash hydrogen-burning luminosity peak.  IR1
is about 0.16 magnitudes below the interflash peak.  In NGC\,419, the
two very dusty AGB stars, IR1 and MIR1, are the most luminous stars in
the cluster.

Pulsation masses have been derived for the dusty stars in each cluster
from nonlinear pulsation models.  The pulsation masses show that these
stars have lost a few tenth of a solar mass due to stellar winds near
the AGB tip.  The AGB tip luminosities are compared to simple
synthetic AGB evolutionary models which include mass loss.  These models
use the derived stellar mass on the early AGB as their starting point.
The evolutionary models which use the Vassiliadis \& Wood (1993) mass
loss rate reproduce the observed tip luminosities quite well while
models using the Bl\"ocker (1995) mass loss rate predict AGB tip
luminosities that are far too faint.

The evolutionary models just described are very approximate.  Full
evolutionary models which include helium shell flashes are required to
properly model the final evolution on the AGB.  The study here shows
that reliable pulsation masses can be determined for AGB stars.  Given
the considerable number of intermediate age clusters in the LMC and
SMC this leads to the possibility of determining the amount of mass
lost in clusters covering a range of AGB tip luminosity, stellar mass
and metallicity.  Comparing these results with evolutionary models
should allow a determination of the mass loss rate as a function of
mass, luminosity and metallicity.

\section*{Acknowledgements}
TL acknowledges funding by Austrian Science Fund FWF through 
project number P20046-N16.

\label{lastpage}

\end{document}